\documentclass[5p]{elsarticle}

\usepackage{lineno,hyperref}
\modulolinenumbers[5]

\journal{Astronomy and Computing}
\usepackage{listings}

\begin{document}
\begin{frontmatter}

\title{%
The ESA Gaia Archive: Data Release 1}

   \author[ESAC]{J. Salgado\corref{mycorrespondingauthor}}
\cortext[mycorrespondingauthor]{Corresponding author}
\ead{jesus.salgado@sciops.esa.int}

   \author[ESAC,VIGO]{J. Gonz\'alez-N\'u\~nez}
   \author[ESAC]{R. Guti\'errez-S\'anchez}
   \author[ESAC]{J.C. Segovia}
   \author[ESAC]{J. Dur\'an}
   \author[ESAC]{J.L. Hern\'andez}
   \author[ESAC]{C. Arviset}

   \address[ESAC]{European Space Astronomy Centre (ESA/ESAC), P.O.Box 78, 28691 Villanueva de la Ca\~nada, Madrid, Spain}
   \address[VIGO]{ETSE Telecomunicaci\'on, Universidade de Vigo, Campus Lagoas-Marcosende, 36310 Vigo, Galicia, Spain}

\begin{abstract}
ESA Gaia mission is producing the more accurate source catalogue in astronomy up to now. That represents
a challenge on the archiving area to make accessible this information to the astronomers in an efficient way.
Also, new astronomical missions have reinforced the change on the development of archives. Archives, as simple
applications to access the data are being evolving into complex data center structures where computing power
services are available for users and data mining tools are integrated into the server side.
In the case of astronomy science that involves the use of big catalogues, as in Gaia (or Euclid to come), the
common ways to work on the data need to be changed to a new paradigm "move code close to the data", what implies
that data mining functionalities are becoming a must to allow the science exploitation.
To enable these capabilities, a TAP+ interface, crossmatch capabilities, full catalogue histograms, serialisation of
intermediate results in cloud resources like VOSpace, etc have been implemented for the Gaia DR1, to enable the
exploitation of these science resources by the community without the bottlenecks on the connection bandwidth.
We present the architecture, infrastructure and tools already available in the Gaia Archive Data Release 1 \\
(\url{http://archives.esac.esa.int/gaia/}) and we describe capabilities and infrastructure.
\end{abstract}


\end{frontmatter}

\section{Introduction}
The ESDC (\cite{2015scop.confE...2A}) (ESAC Science Data Centre), located at ESAC, is the responsible of the design and implementation of the
science astrophysical and planetary missions archives. This group is also responsible of the long term
preservation of the data.

The Gaia satellite (\cite{2016A&A...595A...1G}), launched last 19th December 2013, has produce a stereoscopic census of the galaxy and its first science main product is a catalogue that contains 1.14 billion identified sources. This catalogue, that includes very accurate positions, was released to the scientific community last 14 September 2016 . Also, other transient data is being obtained with some statistics (up to 5th June 2015) of 12TB of science telemetry, 1.5E9 spectra and a full database content of 44TB of data still under reduction process. The final data delivery will not only include the catalogue of one billion sources with accurate positions and proper motions but, also, the single epoch CCD transit data that was used in its computation, making an estimated total around 1 PB at the end of the mission in 2022.

On the other hand and to provide context to the data, other catalogues commonly used in astronomy should be also provided by any archive that exposes Gaia data, in order to provide support to several science use cases like transient data, photometric emission in other bands, accurate proper motions, etc

All this data should be made available in a simple but performant interface that allows the users to explore the Gaia data results and implement scientific use cases. The interface provided by ESA is the ESA Gaia Archive, located at \url{http://archives.esac.esa.int/gaia/}.

This is the first of some space-based astronomical missions that implies big amounts of data to be exposed to the community. Future missions like, e.g. Euclid, will provide even bigger numbers. That forces a change in the paradigm of developing archives and, also, in
the way the scientists interact with the data. In order to manage all these data for science cases, the new archives requires to have the possibility to execute processes close to the data (preventing, as much as possible, internet traffic of mission data).

Usually, scientists have interacted with catalogues by either downloading full catalogues and working in their local space or by doing queries to a REST service. In order to manage big amounts of data from the server to the client, the users tried to select regions of the sky (if the service allows for it) and combine the results in their local disk by, e.g.,
crosmatching the results with their own catalogues. However, this procedure is cumbersome and there was a need for improvement.

IVOA TAP protocol (\cite{2010ivoa.spec.0327D})  specifies how to interact with a big catalogue by using a SQL-like language adapted for astronomy called ADQL (Astronomical Data Query Language (\cite{2008ivoa.spec.1030O})). TAP protocol also identifies specific operations like volatile table upload. That means, a local catalogue is uploaded and a query is done afterwards. This operation is usually repeated several times during a specific use case so it becomes a bottleneck for the users.

For the case of the Gaia archive, the standard TAP protocol has been extended with extra functionalities like private user spaces at the server DB. Tables uploaded are then statically present and available for this user for different operations. Also, tables can be shared to other users so collaborative work is allowed. This concept is called the TAP+ protocol.

Jobs handler during the execution of asynchronous queries is defined in standard TAP by the IVOA Universal Worker Service Pattern (UWS (\cite{2010ivoa.spec.1010H})). This protocol has been also extended by the Gaia Archive through the creation of the UWS+ pattern.

Also, in order to share files, a VOSpace (\cite{2013ivoa.spec.0329G}) instance has been developed and integrated within the TAP+ service so the results of the queries can be stored into the user VOSpace area and shared as files to other users.

\section{Protocols}
As said, standard TAP and UWS recommendations do not specify some capabilities required by the Gaia Archive. For that reason, we have created TAP+ and UWS+, which are extensions of the IVOA specifications that maintain backwards compatibility with the standard protocols, so VO applications can connect transparently to them. New capabilities are:
\begin{itemize}
\item \textbf{Authentication mechanism:} TAP specifies the access to the entire system without restrictions (i.e. all data must be public). TAP+ defines proprietary data with access rights at user level. These access rights also cover the tables sharing mechanism described below.
\item \textbf{User spaces:} TAP specifies how to upload a data table, how to execute a query on that table and retrieve the results. The uploaded table is removed after the query is executed so, if a different operation on this table is needed, the user needs to upload the table again. TAP+ avoids unnecessary data traffic by the creation of user spaces where any user can store tables persistently. Doing that, user tables can be used by the owners for any query at any time. As the tables are stored into a database and as the results are also kept on the server disk until the user decides to remove them. User quotas are defined and managed at user level to increase and decrease them on-demand.
\item \textbf{Tables sharing:} TAP+ supports a mechanism to handle users groups in order to share tables from their user spaces. When a table is shared, the access mechanism authorizes selected users to access to the table without duplicating the table content. By doing that, the shared table consumes disk only at the user space of the owner. The sharing mechanism will generate notifications to the new users or group of users that can access to the shared tables so they became aware of the new shared item.
\item \textbf{Multiple schemas:} TAP protocol defines a service capability called TAP Schema that self-describes tables and columns available in the system. By doing that, TAP clients can discover the structure of the data contained on this TAP service instance and define valid queries. As the Gaia Archive provides access to different users with different privileges, it is necessary to provide different TAP Schema views for different users taking into account user spaces, shared data and privileges.
\item \textbf{Cross-match:} Apart from pre-computed cross-match tables produced to link Gaia sources with some of the main astronomical catalogues, the Gaia Archive provides a cross-match close-neighbours function added to the ADQL grammar. This function is only available for registered users as it implies the creation of an intermediate table in the user private space. By using this function, users can find counterparts of user tables sources and the sources present into other Gaia Archive astronomical catalogues.
\end{itemize}

\section{Data Content}
These are some of the main tables included in the Gaia Archive DR1 version:

\begin{itemize}
\item Gaia main catalogue (Gaia Source): Main Gaia DR1 product that contains all Gaia observed sources.

\item Tycho-Gaia astrometric solution (TGAS): Subset of GaiaSource comprising those stars in the Hipparcos and Tycho-2 Catalogues for which a full 5-parameter astrometric solution has been possible in Gaia Data Release 1.

\item External catalogues: Some of the main catalogues in astronomy are also part of the Gaia archive to allow science use cases. Catalogues included are
AllWISE (\cite{2011ApJ...731...53M}),
UCAC4 (\cite{2013AJ....145...44Z}),
SDSS DR9 (\cite{2012ApJS..203...21A}),
2MASS Point Source Catalogue (\cite{2006AJ....131.1163S}),
URAT1 (\cite{2015AAS...22543301Z}),
GSC23 (\cite{2008AJ....136..735L}) and
PPMXL (\cite{2010AJ....139.2440R}).

\item Precomputed crossmatch tables: All the external catalogues mentioned in previous point have been crossmatched with the Gaia Source DR1 table, creating two tables per catalogue: one with the best candidate of the crossmatch (best neighbour) and another reflecting the neighbourhood density.

\item gaiadr1.variable\_summary: Table of sources tagged as variables in the Gaia Source table.

\item gaiadr1.phot\_variable\_time\_series\_gfov: Field-of-view time series of sources tagged as variables in the Gaia Source table.

\item gaiadr1.rrlyrae: RR Lyrae stars identified in the Variable Summary table.

\item gaiadr1.cepheid: Cepheid stars identified in the Variable Summary table.

\item gaiadr1.ext\_phot\_zero\_point: Definition of the Gaia photometric system: for GDR1 only zero-points will be computed, one for G and one for each integrated BP and integrated RP.

\end{itemize}

\section{Archive interfaces}
As described previously, the backend of the Archive consists on a TAP+ server. This server listens for HTTP requests as defined by the TAP protocol, on top of which extended capabilities have been developed. The Gaia Archive provides access to the Gaia data through a Graphical User Interface (GUI) and a command line interface. For the second one, a Java and a python example clients are also provided. However, the use of a standard IVOA protocols allows the access to the data through the usage of a big set of VO-compatible clients.

\subsection{Graphical User Interface (GUI)}

The Gaia Archive GUI is a web AJAX application that provides easy access to the functionalities of the TAP+ server. The use of web technologies maximise the accessibility in several ways. To start with, no specific software has to be installed in the client machine. The only software requisite to use the application is a web browser. This also increases the inter-platform compatibility, extending the usage from desktop or laptop PCs to mobile devices.

The main functions provided by the interface are described in the following sections.

\begin{figure}
  \resizebox{\hsize}{!}{\includegraphics{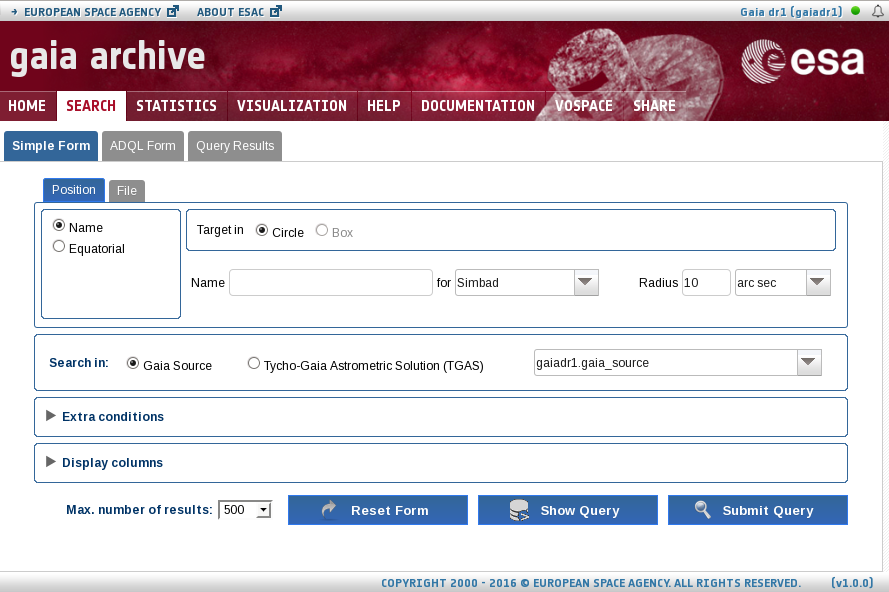}}
  \caption{Simple query form}
  \label{FigSimpleForm}
\end{figure}

\subsubsection{Search}

The search interface provides access to the main functionality of the archive. The user may choose between two search interfaces:

\begin{description}
\item [Simple Form] Intended to be a simple tool to start exploring the archive. It provides name resolution for object names and the possibility of tailoring a query that can be either executed to browse results or sent to the more advanced and flexible ADQL form.

\item [ADQL form] In this interface the user is able to create complex queries in ADQL format with a very rich astronomy oriented querying languaje. The list of available tables is provided, comprising publicly accessible tables, tables in the private user area and tables shared to the user by other users of the system.

\begin{figure}
  \resizebox{\hsize}{!}{\includegraphics{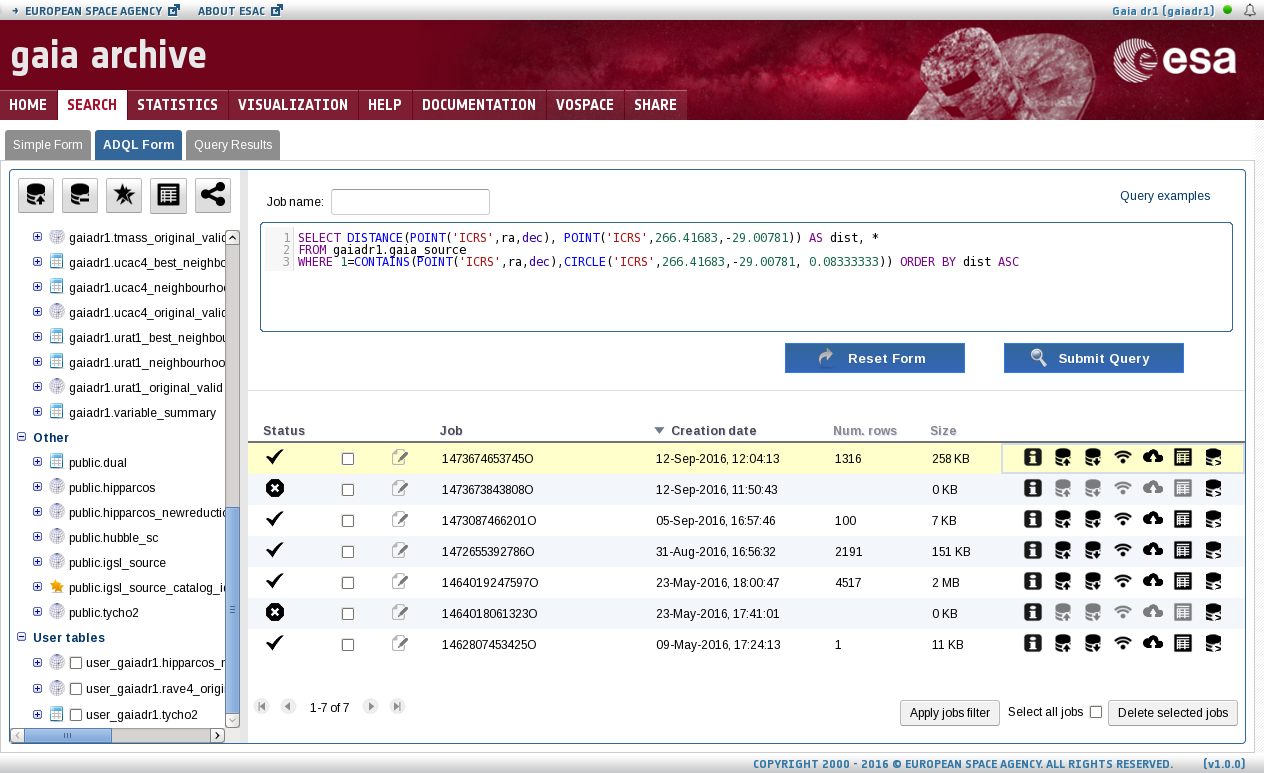}}
  \caption{Advanced ADQL query form. Left: available tables. Left top: table related functions (upload, delete, cross-match, edit and share). Top: ADQL query editor. Bottom: job managemente window}
  \label{FigAdqlForm}
\end{figure}

The execution of an ADQL query produces a job which status can be monitored in a interactive panel. Once the job is finished, it can be browsed, downloaded, uploaded to the user's VOSpace area or sent to other VO-enabled applications using the SAMP protocol.
\end{description}

\subsubsection{Statistics}
The archive provides precomputed statistical information of the main catalogs. This comprises density maps for the whole population and histograms for the individual columns.

\begin{figure}
  \resizebox{\hsize}{!}{\includegraphics{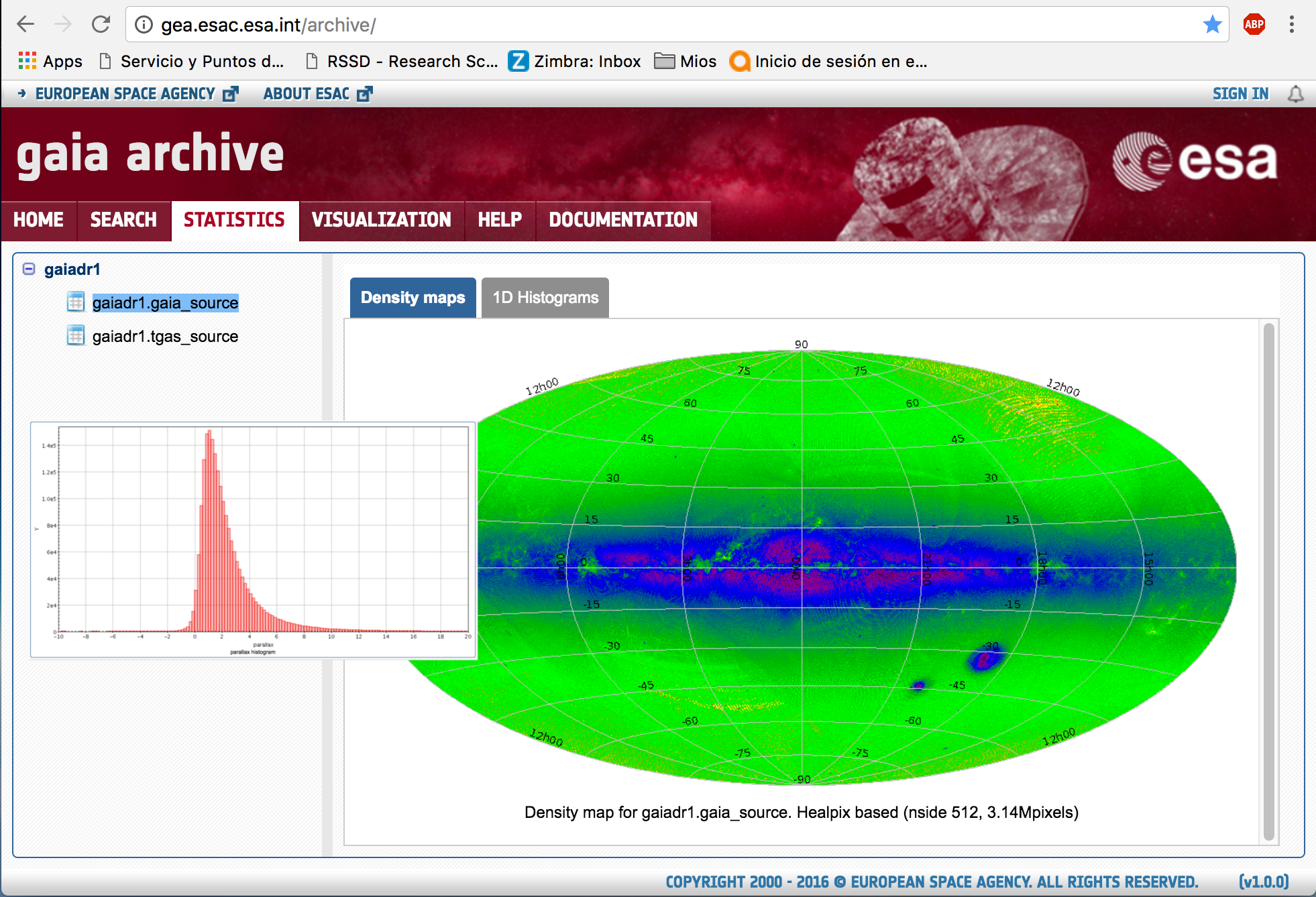}}
  \caption{Pre-computed density maps and histograms for main Gaia catalog.}
  \label{FigStatistics}
\end{figure}

\subsubsection{Sharing capabilities}
The Gaia Archive have been developed to meet the needs of an increasingly cooperative scientific community. For this purpose, the archive have been designed with capabilities that enable users to share data. Intermediate or final results of a research can be shared between a workgroup without the need of transferring or duplicating data. Any table inside the user private schema can be shared with one or more groups. These groups can be created and maintained by the user itself. Shared tables can be used in queries by the authorized users. Size of the shared tables will only be accounted for the owner space quota.

\subsection{Command line interface}
Graphical User Interface provides a fast and easy way to interact with the Gaia catalogs. However, efficient techniques for analysing big amounts of data implies the use of automated programatic access to the archive capabilities. Gaia TAP+ provides a REST interface that can be easily used from scripts. The archive help gathers a set of examples where the use of the archive capabilities from a command line interface is shown. It is quite straightforward to construct complex command line scripts from the individual calls explained in that section. All the functionality at server side used by the User Interface has been documented into the Gaia Archive and can be invoked through the command line interface.

\section{Infrastructure}

The Gaia Archive infrastructure has been designed with the following principles:
\begin{itemize}
\item \textbf{High throughput:} All components of the system have been individually tested for performance; incremental load and stress testing has been applied to the overall system in 4 incremental iterations identifying performance bottlenecks, reverting back to the system as improvements for next iteration.
\item \textbf{High availability:} Through warm standby and redundancy of critical components, the Gaia Archive infrastructure can reduce recovery times from a major malfunction through a switch to an alternate component.
\end{itemize}

\begin{figure}
  \resizebox{\hsize}{!}{\includegraphics{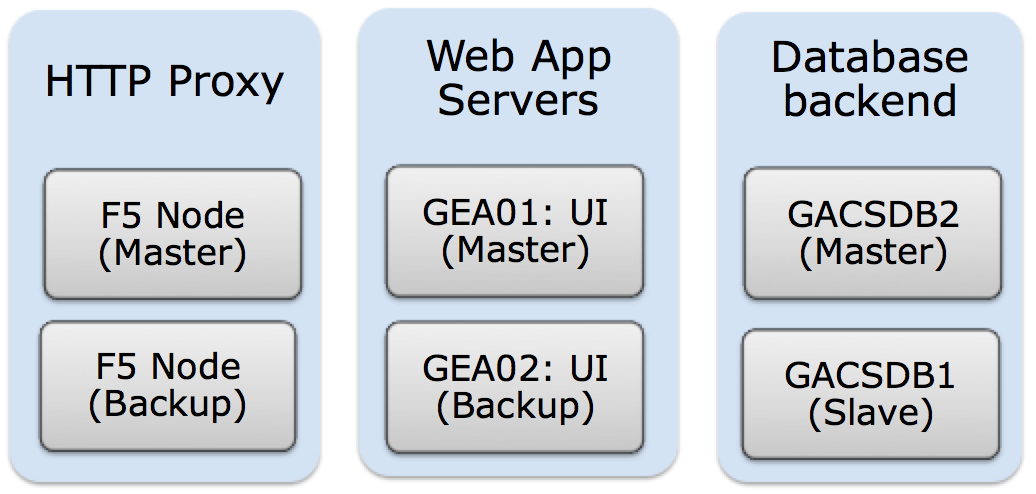}}
  \caption{Gaia Archive Infrastructure}
  \label{FigInfrastructure}
\end{figure}

Figure \ref{FigInfrastructure} provides an overview of how the redundant components contribute to a two tiered architecture (Web + Database) with a front HTTP proxy layer. In the DR1 configuration, only F5(Master), \textbf{GEA01} and \textbf{GACSDB02} receive traffic, with backup nodes being in a warm standby configuration, ready to take over in case of systems failure.

\subsection{Frontend}
The archive front-end layer is deployed in two web application server optimised machines, with similar specifications \textbf{GEA01} and \textbf{GEA02} (2 X Xeon E5-2670 (2x12 cores total) and 64 GB of RAM)

\subsection{Databases}

Database subsystem is based in a PostgreSQL 9.5 installation with pgSphere
(\cite{2004ASPC..314..225C}) and Q3C (\cite{2006ASPC..351..735K}) contribution modules. In the current version of the Archive (1.0), all geometrical queries are handled through Q3C indexing.

Warm standby is achieved through the set up of Streaming Replication from the master machine \textbf{GACSDB02} (4 X Xeon E7-4850 (4x12 cores total), 1514GB of RAM) to the slave machine \textbf{GACSDB01} (4 X Xeon E5-4640 (4x8 cores total), 1009 GB of RAM).

Under a nominal scenario all traffic is handled through \textbf{GACSDB02}. Upon failure of this machine, \textbf{GACSDB01} is promoted to master and all traffic handled through it.

Storage for the database system is layered into three sections, with decreasing level of throughput in terms of IOPs:
\begin{description}
\item [\textbf{Level 1:}] Gaia DR1 catalogues/tables.
\item [\textbf{Level 2:}] External catalogues and crossmatch tables.
\item [\textbf{Level 3:}] User spaces.
\end{description}

In the master database server for DR1, GACSDB02, both Level 1 and Level 2 are stored in local SSD storage. In the slave standby server, Level 1 data is stored in PCI-e SSD flash disk, and Level 2 data in local SAS disks.

Level 3 data is stored in network attached storage volumes for all databases, to allow for scaling up volumes as user uploaded tables increase the overall volume.

\subsection{File Server}
For the bulk download of full Gaia catalogues, a file download server has been setup through the use of a Content Delivery Network (CDN). This infrastructure brings a high level of scalability through the replication of Gaia data in a distributed network of proxy servers deployed in multiple data centres dedicated to different world regions. It also offloads most heavy traffic from the Gaia Archive components at ESAC and from the overall ESAC network connectivity.

This CDN operates during the first phase of the DR1 publication and can be reverted back to a file server set up at ESAC once the traffic peak of the DR1 is over.

\subsection{Network}
All network components and links between the different system machines, including connectivity with network attached storage volumes has been designed and tested to deliver 10Gbps.

\section{Conclusions}
The Gaia Archive provides the protocols and scalability to allow easy access to the big catalogues generated by the Gaia mission during DR1. It reuses and extends standard inteoperability protocols defined within the IVOA so it allows users the access to these data through well documented interfaces and VO tools.

Also, we describe the needed infrastructure created at ESAC to guarantee the execution of complex use cases at server side, based on the move code close to the
data paradigm.

This approach will be reused and extended for future archives of ESA astronomical missions.


\begin{thebibliography}{10}
\expandafter\ifx\csname url\endcsname\relax
  \def\url#1{\texttt{#1}}\fi
\expandafter\ifx\csname urlprefix\endcsname\relax\def\urlprefix{URL }\fi
\expandafter\ifx\csname href\endcsname\relax
  \def\href#1#2{#2} \def\path#1{#1}\fi

\bibitem{2015scop.confE...2A}
C.~{Arviset}, {Science Archives at the ESAC Science Data Centre}, in: Science
  Operations 2015: Science Data Management, id.2, 2015, p.~2.
\newblock \href {http://dx.doi.org/10.5281/zenodo.34493}
  {\path{doi:10.5281/zenodo.34493}}.

\bibitem{2016A&A...595A...2G}
{Gaia Collaboration}, A.~G.~A. {Brown}, A.~{Vallenari}, T.~{Prusti}, J.~H.~J.
  {de Bruijne}, F.~{Mignard}, R.~{Drimmel}, C.~{Babusiaux}, C.~A.~L.
  {Bailer-Jones}, U.~{Bastian}, et~al., {Gaia Data Release 1. Summary of the
  astrometric, photometric, and survey properties} 595 (2016) A2.
\newblock \href {http://arxiv.org/abs/1609.04172} {\path{arXiv:1609.04172}},
  \href {http://dx.doi.org/10.1051/0004-6361/201629512}
  {\path{doi:10.1051/0004-6361/201629512}}.

\bibitem{2016A&A...595A...1G}
{Gaia Collaboration}, T.~{Prusti}, J.~H.~J. {de Bruijne}, A.~G.~A. {Brown},
  A.~{Vallenari}, C.~{Babusiaux}, C.~A.~L. {Bailer-Jones}, U.~{Bastian},
  M.~{Biermann}, D.~W. {Evans}, et~al., {The Gaia mission} 595 (2016) A1.
\newblock \href {http://arxiv.org/abs/1609.04153} {\path{arXiv:1609.04153}},
  \href {http://dx.doi.org/10.1051/0004-6361/201629272}
  {\path{doi:10.1051/0004-6361/201629272}}.

\bibitem{RFC6690}
Z.~Shelby, \href{http://www.rfc-editor.org/rfc/rfc6690.txt}{Constrained restful
  environments (core) link format}, RFC 6690, RFC Editor,
  \url{http://www.rfc-editor.org/rfc/rfc6690.txt} (August 2012).
\newline\urlprefix\url{http://www.rfc-editor.org/rfc/rfc6690.txt}

\bibitem{2010ivoa.spec.0327D}
P.~{Dowler}, G.~{Rixon}, D.~{Tody}, {Table Access Protocol Version 1.0}, IVOA
  Recommendation 27 March 2010 (Mar. 2010).
\newblock \href {http://arxiv.org/abs/1110.0497} {\path{arXiv:1110.0497}}.

\bibitem{2008ivoa.spec.1030O}
P.~{Osuna}, I.~{Ortiz}, J.~{Lusted}, P.~{Dowler}, A.~{Szalay}, Y.~{Shirasaki},
  M.~A. {Nieto-Santisteban}, M.~{Ohishi}, W.~{O'Mullane}, {VOQL-TEG Group},
  {VOQL Working Group.}, {IVOA Astronomical Data Query Language Version 2.00},
  IVOA Recommendation 30 October 2008 (Oct. 2008).
\newblock \href {http://arxiv.org/abs/1110.0503} {\path{arXiv:1110.0503}}.

\bibitem{2005cs........2072O}
W.~{OMullane}, N.~{Li}, M.~{Nieto-Santisteban}, A.~{Szalay}, A.~{Thakar},
  J.~{Gray}, {Batch is back: CasJobs, serving multi-TB data on the Web}\href
  {http://arxiv.org/abs/cs/0502072} {\path{arXiv:cs/0502072}}.

\bibitem{2010ivoa.spec.1010H}
P.~{Harrison}, G.~{Rixon}, {Universal Worker Service Pattern Version 1.0}, IVOA
  Recommendation 10 October 2010 (Oct. 2010).
\newblock \href {http://arxiv.org/abs/1110.0510} {\path{arXiv:1110.0510}}.

\bibitem{2013ivoa.spec.0329G}
M.~{Graham}, D.~{Morris}, G.~{Rixon}, P.~{Dowler}, A.~{Schaaff}, D.~{Tody},
  {VOSpace specification Version 2.0}, IVOA Recommendation 29 March 2013 (Mar.
  2013).
\newblock \href {http://arxiv.org/abs/1509.06049} {\path{arXiv:1509.06049}}.

\bibitem{2015A&A...574A.115M}
D.~{Michalik}, L.~{Lindegren}, D.~{Hobbs}, {The Tycho-Gaia astrometric solution
  . How to get 2.5 million parallaxes with less than one year of Gaia data} 574
  (2015) A115.
\newblock \href {http://arxiv.org/abs/1412.8770} {\path{arXiv:1412.8770}},
  \href {http://dx.doi.org/10.1051/0004-6361/201425310}
  {\path{doi:10.1051/0004-6361/201425310}}.

\bibitem{2011ApJ...731...53M}
A.~{Mainzer}, J.~{Bauer}, T.~{Grav}, J.~{Masiero}, R.~M. {Cutri}, J.~{Dailey},
  P.~{Eisenhardt}, R.~S. {McMillan}, E.~{Wright}, R.~{Walker}, R.~{Jedicke},
  T.~{Spahr}, D.~{Tholen}, R.~{Alles}, R.~{Beck}, H.~{Brandenburg},
  T.~{Conrow}, T.~{Evans}, J.~{Fowler}, T.~{Jarrett}, K.~{Marsh}, F.~{Masci},
  H.~{McCallon}, S.~{Wheelock}, M.~{Wittman}, P.~{Wyatt}, E.~{DeBaun},
  G.~{Elliott}, D.~{Elsbury}, T.~{Gautier}, IV, S.~{Gomillion}, D.~{Leisawitz},
  C.~{Maleszewski}, M.~{Micheli}, A.~{Wilkins}, {Preliminary Results from
  NEOWISE: An Enhancement to the Wide-field Infrared Survey Explorer for Solar
  System Science} 731 (2011) 53.
\newblock \href {http://arxiv.org/abs/1102.1996} {\path{arXiv:1102.1996}},
  \href {http://dx.doi.org/10.1088/0004-637X/731/1/53}
  {\path{doi:10.1088/0004-637X/731/1/53}}.

\bibitem{2013AJ....145...44Z}
N.~{Zacharias}, C.~T. {Finch}, T.~M. {Girard}, A.~{Henden}, J.~L. {Bartlett},
  D.~G. {Monet}, M.~I. {Zacharias}, {The Fourth US Naval Observatory CCD
  Astrograph Catalog (UCAC4)} 145 (2013) 44.
\newblock \href {http://arxiv.org/abs/1212.6182} {\path{arXiv:1212.6182}},
  \href {http://dx.doi.org/10.1088/0004-6256/145/2/44}
  {\path{doi:10.1088/0004-6256/145/2/44}}.

\bibitem{2012ApJS..203...21A}
C.~P. {Ahn}, R.~{Alexandroff}, C.~{Allende Prieto}, S.~F. {Anderson},
  T.~{Anderton}, B.~H. {Andrews}, {\'E}.~{Aubourg}, S.~{Bailey}, E.~{Balbinot},
  R.~{Barnes}, et~al., {The Ninth Data Release of the Sloan Digital Sky Survey:
  First Spectroscopic Data from the SDSS-III Baryon Oscillation Spectroscopic
  Survey} 203 (2012) 21.
\newblock \href {http://arxiv.org/abs/1207.7137} {\path{arXiv:1207.7137}},
  \href {http://dx.doi.org/10.1088/0067-0049/203/2/21}
  {\path{doi:10.1088/0067-0049/203/2/21}}.

\bibitem{2006AJ....131.1163S}
M.~F. {Skrutskie}, R.~M. {Cutri}, R.~{Stiening}, M.~D. {Weinberg},
  S.~{Schneider}, J.~M. {Carpenter}, C.~{Beichman}, R.~{Capps}, T.~{Chester},
  J.~{Elias}, J.~{Huchra}, J.~{Liebert}, C.~{Lonsdale}, D.~G. {Monet},
  S.~{Price}, P.~{Seitzer}, T.~{Jarrett}, J.~D. {Kirkpatrick}, J.~E. {Gizis},
  E.~{Howard}, T.~{Evans}, J.~{Fowler}, L.~{Fullmer}, R.~{Hurt}, R.~{Light},
  E.~L. {Kopan}, K.~A. {Marsh}, H.~L. {McCallon}, R.~{Tam}, S.~{Van Dyk},
  S.~{Wheelock}, {The Two Micron All Sky Survey (2MASS)} 131 (2006) 1163--1183.
\newblock \href {http://dx.doi.org/10.1086/498708} {\path{doi:10.1086/498708}}.

\bibitem{2015AAS...22543301Z}
N.~{Zacharias}, C.~T. {Finch}, J.~P. {Subasavage}, T.~{Tilleman},
  M.~{DiVittorio}, H.~C. {Harris}, T.~{Rafferty}, G.~{Wieder}, E.~{Ferguson},
  C.~{Kilian}, A.~{Rhodes}, M.~{Schultheis}, {The U.S. Naval Observatory
  Robotic Astrometric Telescope 1st Catalog (URAT1)}, in: American Astronomical
  Society Meeting Abstracts, Vol. 225 of American Astronomical Society Meeting
  Abstracts, 2015, p. 433.01.

\bibitem{2008AJ....136..735L}
B.~M. {Lasker}, M.~G. {Lattanzi}, B.~J. {McLean}, B.~{Bucciarelli},
  R.~{Drimmel}, J.~{Garcia}, G.~{Greene}, F.~{Guglielmetti}, C.~{Hanley},
  G.~{Hawkins}, C.~{Laidler}, V.~G.and~{Loomis}, M.~{Meakes}, R.~{Mignani},
  R.~{Morbidelli}, J.~{Morrison}, R.~{Pannunzio}, A.~{Rosenberg}, M.~{Sarasso},
  R.~L. {Smart}, A.~{Spagna}, C.~R. {Sturch}, A.~{Volpicelli}, R.~L. {White},
  D.~{Wolfe}, A.~{Zacchei}, {The Second-Generation Guide Star Catalog:
  Description and Properties} 136 (2008) 735--766.
\newblock \href {http://arxiv.org/abs/0807.2522} {\path{arXiv:0807.2522}},
  \href {http://dx.doi.org/10.1088/0004-6256/136/2/735}
  {\path{doi:10.1088/0004-6256/136/2/735}}.

\bibitem{2010AJ....139.2440R}
S.~{Roeser}, M.~{Demleitner}, E.~{Schilbach}, {The PPMXL Catalog of Positions
  and Proper Motions on the ICRS. Combining USNO-B1.0 and the Two Micron All
  Sky Survey (2MASS)} 139 (2010) 2440--2447.
\newblock \href {http://arxiv.org/abs/1003.5852} {\path{arXiv:1003.5852}},
  \href {http://dx.doi.org/10.1088/0004-6256/139/6/2440}
  {\path{doi:10.1088/0004-6256/139/6/2440}}.

\bibitem{2017A&A.Marrese}
P.~{Marrese}, S.~{Marinoni}, G.~{Giuffrida}, M.~{Fabrizio}, {Gaia Data Release
  1. Cross-match with external catalogues: algorithm and statistics, A\&A,
  submitted}.

\bibitem{GENIUS}
X.~{Luri}, {Gaia European Network for Improved User Services
  (GENIUS),\url{https://gaia.ub.edu/Twiki/pub/GENIUS/ReferenceDocuments/GENIUS.pdf}}
  (2013).

\bibitem{2009ivoa.spec.0421B}
T.~{Boch}, M.~{Fitzpatrick}, M.~{Taylor}, A.~{Allan}, L.~{Paioro}, J.~{Taylor},
  D.~{Tody}, {SAMP Simple Application Messaging Protocol Version 1.11}, IVOA
  Recommendation 21 April 2009 (Apr. 2009).

\bibitem{RFC4519}
A.~Sciberras, \href{http://www.rfc-editor.org/rfc/rfc4519.txt}{Lightweight
  directory access protocol (ldap): Schema for user applications}, RFC 4519,
  RFC Editor, \url{http://www.rfc-editor.org/rfc/rfc4519.txt} (June 2006).
\newline\urlprefix\url{http://www.rfc-editor.org/rfc/rfc4519.txt}

\bibitem{2016arXiv161003159M}
D.~{Muna}, M.~{Alexander}, A.~{Allen}, R.~{Ashley}, D.~{Asmus}, R.~{Azzollini},
  M.~{Bannister}, R.~{Beaton}, A.~{Benson}, G.~B. {Berriman}, M.~{Bilicki},
  P.~{Boyce}, J.~{Bridge}, J.~{Cami}, E.~{Cangi}, X.~{Chen}, N.~{Christiny},
  C.~{Clark}, M.~{Collins}, J.~{Comparat}, N.~{Cook}, D.~{Croton}, I.~{Delberth
  Davids}, {\'E}.~{Depagne}, J.~{Donor}, L.~A. {dos Santos}, S.~{Douglas},
  A.~{Du}, M.~{Durbin}, D.~{Erb}, D.~{Faes}, J.~G. {Fern{\'a}ndez-Trincado},
  A.~{Foley}, S.~{Fotopoulou}, S.~{Frimann}, P.~{Frinchaboy}, R.~{Garcia-Dias},
  A.~{Gawryszczak}, E.~{George}, S.~{Gonzalez}, K.~{Gordon}, N.~{Gorgone},
  C.~{Gosmeyer}, K.~{Grasha}, P.~{Greenfield}, R.~{Grellmann}, J.~{Guillochon},
  M.~{Gurwell}, M.~{Haas}, A.~{Hagen}, D.~{Haggard}, T.~{Haines}, P.~{Hall},
  W.~{Hellwing}, E.~C. {Herenz}, S.~{Hinton}, R.~{Hlozek}, J.~{Hoffman},
  D.~{Holman}, B.~W. {Holwerda}, A.~{Horton}, C.~{Hummels}, D.~{Jacobs},
  J.~{Juel Jensen}, D.~{Jones}, A.~{Karick}, L.~{Kelley}, M.~{Kenworthy},
  B.~{Kitchener}, D.~{Klaes}, S.~{Kohn}, P.~{Konorski}, C.~{Krawczyk},
  K.~{Kuehn}, T.~{Kuutma}, M.~T. {Lam}, R.~{Lane}, J.~{Liske},
  D.~{Lopez-Camara}, K.~{Mack}, S.~{Mangham}, Q.~{Mao}, D.~J.~E. {Marsh},
  C.~{Mateu}, L.~{Maurin}, J.~{McCormac}, I.~{Momcheva}, H.~{Monteiro},
  M.~{Mueller}, R.~{Munoz}, R.~{Naidu}, N.~{Nelson}, C.~{Nitschelm},
  C.~{North}, J.~{Nunez-Iglesias}, S.~{Ogaz}, R.~{Owen}, J.~{Parejko},
  V.~{Patr{\'{\i}}cio}, J.~{Pepper}, M.~{Perrin}, T.~{Pickering},
  J.~{Piscionere}, R.~{Pogge}, R.~{Poleski}, A.~{Pourtsidou}, A.~M.
  {Price-Whelan}, M.~L. {Rawls}, S.~{Read}, G.~{Rees}, H.~{Rein}, T.~{Rice},
  S.~{Riemer-S{\o}rensen}, N.~{Rusomarov}, S.~F. {Sanchez},
  M.~{Santander-Garc{\'{\i}}a}, G.~{Sarid}, W.~{Schoenell}, A.~{Scholz}, R.~L.
  {Schuhmann}, W.~{Schuster}, P.~{Scicluna}, M.~{Seidel}, L.~{Shao},
  P.~{Sharma}, A.~{Shulevski}, D.~{Shupe}, C.~{Sif{\'o}n}, B.~{Simmons},
  M.~{Sinha}, I.~{Skillen}, B.~{Soergel}, T.~{Spriggs}, S.~{Srinivasan},
  A.~{Stevens}, O.~{Streicher}, E.~{Suchyta}, J.~{Tan}, O.~G. {Telford},
  R.~{Thomas}, C.~{Tonini}, G.~{Tremblay}, S.~{Tuttle}, T.~{Urrutia},
  S.~{Vaughan}, M.~{Verdugo}, A.~{Wagner}, J.~{Walawender}, A.~{Wetzel},
  K.~{Willett}, P.~K.~G. {Williams}, G.~{Yang}, G.~{Zhu}, A.~{Zonca}, {The
  Astropy Problem}\href {http://arxiv.org/abs/1610.03159}
  {\path{arXiv:1610.03159}}.

\bibitem{2004ASPC..314..225C}
I.~{Chilingarian}, O.~{Bartunov}, J.~{Richter}, T.~{Sigaev}, {PostgreSQL: the
  Suitable DBMS Solution for Astronomy and Astrophysics}, in: F.~{Ochsenbein},
  M.~G. {Allen}, D.~{Egret} (Eds.), Astronomical Data Analysis Software and
  Systems (ADASS) XIII, Vol. 314 of Astronomical Society of the Pacific
  Conference Series, 2004, p. 225.

\bibitem{2006ASPC..351..735K}
S.~{Koposov}, O.~{Bartunov}, {Q3C, Quad Tree Cube -- The new Sky-indexing
  Concept for Huge Astronomical Catalogues and its Realization for Main
  Astronomical Queries (Cone Search and Xmatch) in Open Source Database
  PostgreSQL}, in: C.~{Gabriel}, C.~{Arviset}, D.~{Ponz}, S.~{Enrique} (Eds.),
  Astronomical Data Analysis Software and Systems XV, Vol. 351 of Astronomical
  Society of the Pacific Conference Series, 2006, p. 735.

\bibitem{2017A&C....20...77G}
J.~{Gonz{\'a}lez-N{\'u}{\~n}ez}, R.~{Guti{\'e}rrez-S{\'a}nchez}, J.~{Salgado},
  J.~C. {Segovia}, B.~{Mer{\'{\i}}n}, F.~{Aguado-Agelet}, {A parallel model for
  SQL astronomical databases based on solid state storage. Application to the
  Gaia Archive PostgreSQL database}, Astronomy and Computing 20 (2017) 77--82.
\newblock \href {http://dx.doi.org/10.1016/j.ascom.2017.03.006}
  {\path{doi:10.1016/j.ascom.2017.03.006}}.

\end{thebibliography}
\end{document}